\documentclass[aps,floatfix,prl]{revtex4}
\usepackage[latin9,utf8]{inputenc}
\usepackage{float}
\usepackage{textcomp}
\usepackage{amsmath}
\usepackage{graphicx}
\makeatletter
\usepackage{url}
\usepackage{grffile}
\usepackage{bbold}

\newcommand{\lyxmathsym}[1]{\ifmmode\begingroup\def\b@ld{bold}
  \text{\ifx\math@version\b@ld\bfseries\fi#1}\endgroup\else#1\fi}

\@ifundefined{textcolor}{}
{%
 \definecolor{BLACK}{gray}{0}
 \definecolor{WHITE}{gray}{1}
 \definecolor{RED}{rgb}{1,0,0}
 \definecolor{GREEN}{rgb}{0,1,0}
 \definecolor{BLUE}{rgb}{0,0,1}
 \definecolor{CYAN}{cmyk}{1,0,0,0}
 \definecolor{MAGENTA}{cmyk}{0,1,0,0}
 \definecolor{YELLOW}{cmyk}{0,0,1,0}
}

\usepackage{dcolumn}
\usepackage{color}
\DeclareGraphicsExtensions{.png .jpg .pdf}

\usepackage{hyperref}
\hypersetup{
     colorlinks = true,
     linkcolor = blue,
     anchorcolor = blue,
     citecolor = blue,
     filecolor = blue,
     urlcolor = blue
     }
\usepackage{braket}
\usepackage{physics}
\usepackage{array}
\usepackage{booktabs}
\usepackage{soul}

\makeatother

\usepackage[printwatermark]{xwatermark}
\usepackage{xcolor}
\usepackage{graphicx}
\usepackage{tikz}

\newsavebox\mybox
\savebox\mybox{\tikz[color=gray,opacity=0.4]\node{arXiv Version};}
\newwatermark*[
  allpages,
  angle=45,
  scale=12,
  xpos=-100,
  ypos=-120
]{\usebox\mybox}

\begin{document}

\title{Zitterbewegung of moir\'e excitons in twisted MoS$_2$/WSe$_2$ hetero-bilayers}

\author{I. R. Lavor}
\affiliation{Universidade Federal do Cear\'a, Departamento de
F\'{\i}sica, 60455-760 Fortaleza, Cear\'a, Brazil}
\affiliation{Instituto Federal de Educação, Ciência e Tecnologia do Maranhão, KM-04, Enseada, 65200-000, Pinheiro, Maranhão, Brazil}
\affiliation{Department of Physics, University of Antwerp, Groenenborgerlaan 171, B-2020 Antwerp, Belgium}
\author{D. R. da Costa}
\affiliation{Universidade Federal do Cear\'a, Departamento de
F\'{\i}sica, 60455-760 Fortaleza, Cear\'a, Brazil}

\author{L. Covaci}
\affiliation{Department of Physics, University of Antwerp, Groenenborgerlaan 171, B-2020 Antwerp, Belgium}

\author{M. V. Milo\v{s}evi\'c}
\affiliation{Department of Physics, University of Antwerp, Groenenborgerlaan 171, B-2020 Antwerp, Belgium}

\author{F. M. Peeters}
\affiliation{Department of Physics, University of Antwerp, Groenenborgerlaan 171, B-2020 Antwerp, Belgium}

\author{A. Chaves} 
\affiliation{Universidade Federal do Cear\'a, Departamento de
F\'{\i}sica, 60455-760 Fortaleza, Cear\'a, Brazil}
\affiliation{Department of Physics, University of Antwerp, Groenenborgerlaan 171, B-2020 Antwerp, Belgium}

\begin{abstract}
The moir\'e pattern observed in stacked non-commensurate crystal lattices, such as hetero-bilayers of transition metal dichalcogenides, produces a periodic modulation of their bandgap. Excitons subjected to this potential landscape exhibit a band structure that gives rise to a quasi-particle dubbed moir\'e exciton. In the case of MoS$_2$/WSe$_2$ hetero-bilayers, the moir\'e trapping potential has honeycomb symmetry and, consequently, the moir\'e exciton band structure is the same as that of a Dirac-Weyl fermion, whose mass can be further tuned down to zero with a perpendicularly applied field. Here we show that, analogously to other Dirac-like particles, moir\'e exciton exhibits a trembling motion, also known as zitterbewegung, whose long timescales are compatible with current experimental techniques for exciton dynamics. This promotes the study of the dynamics of moir\'e excitons in van der Waals heterostructures as an advantageous solid-state platform to probe zitterbewegung, broadly tunable by gating and inter-layer twist angle. 
\end{abstract}

\maketitle
%

\textit{Introduction} Zitterbewegung (ZBW) is a fast trembling motion of elementary particles that obey the Dirac equation~\citep{Dirac1928}, predicted by Erwin Schr{\"o}dinger in $1930$ for relativistic fermions~\citep{Schroedinger1930a}. Schr{\"o}dinger observed that the components of the relativistic velocity for particles in vacuum does not commute with the free-particle Hamiltonian~\citep{Schroedinger1930a}. As a consequence, the expectation value of the position operator for a fermion wave packet displays rapid oscillatory motion, owing to the fact that the velocity is not a constant of motion, as well as to the interference between the positive and negative energy states composing the wave packet~\cite{Schroedinger1930a,greiner2000relativistic, huang1952zitterbewegung}.

\begin{figure}[!t]
\centering{\includegraphics[width=0.5\textwidth]{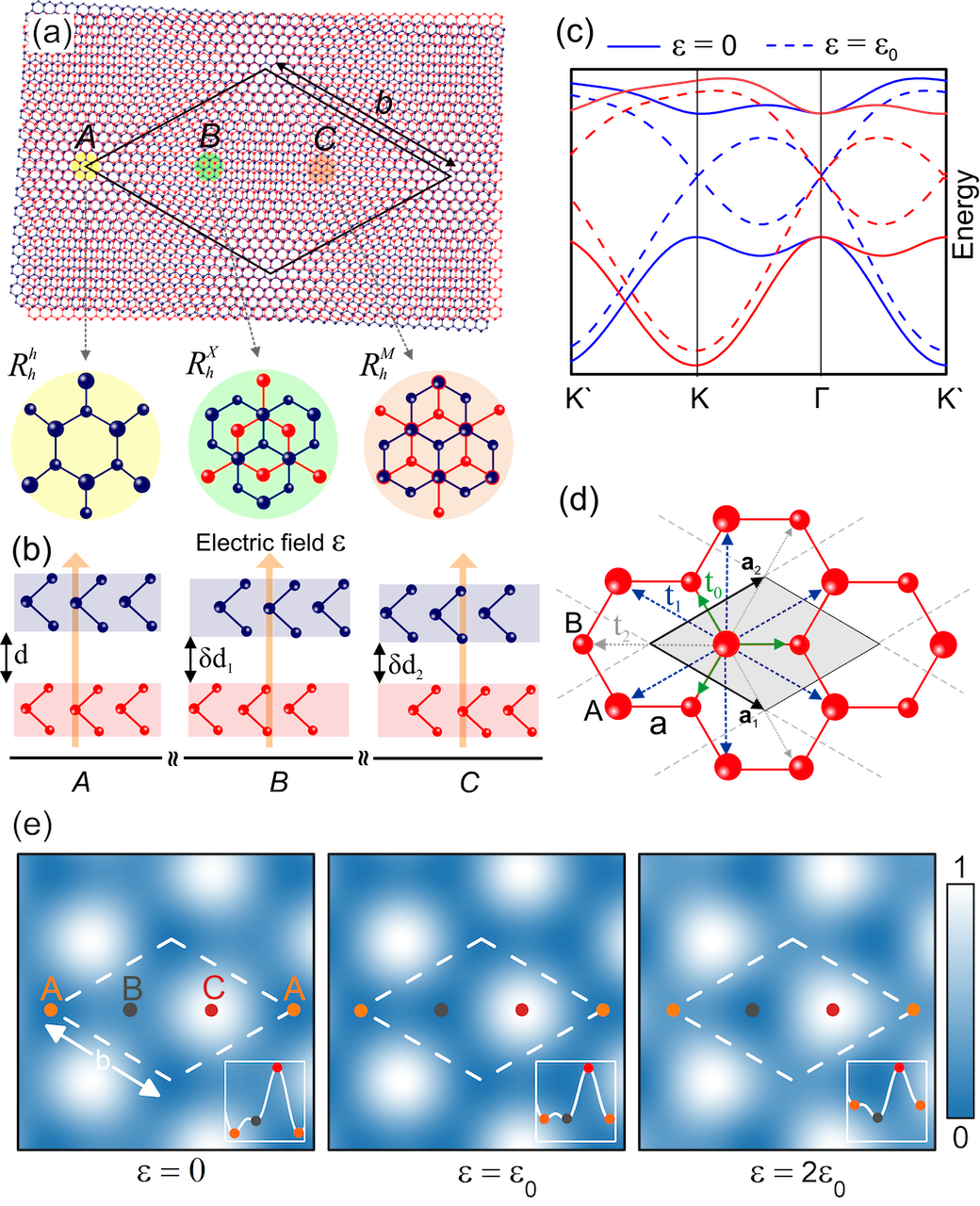}}
    \caption{(a) Moir\'e pattern with period $b$ in an $\text{MoS}_{2}/\text{WSe}_{2}$ hetero-bilayer, twisted by 3$^{\circ}$. Black diamond represents the supercell. Insets magnify three characteristic locations ($A$, $B$ and $C$), where atomic registries resemble lattice-matched bilayers of different R-type stacking. (b) Lateral view of the inter-layer distance of the regions A, B and C (for more details, see Ref.~\citep{yu2017moire}). (c) Corresponding band structures, calculated with the tight-binding model, for the first moir\'e Brillouin zone with (dashed lines) and without (solid lines) an applied electric field ($\varepsilon=\varepsilon_0 \approx0.44~$V/nm) for the $K$ (red) and $K^\prime$ (blue) valleys of the crystal. (d) Representation of a honeycomb lattice structure and unit cell (gray region) where sub-lattice sites $A$ and $B$ correspond to the respective stacking registries labelled in (a), and with lattice constant $a$. First, second and third nearest-neighbors hopping parameters are represented by $t_0$ (green arrows), $t_1$ (blue arrows) and $t_2$ (gray arrows), respectively. $\vec{a}_1$ and $\vec{a}_2$ are the basis vectors. (e) Colormap of the ILE potential landscape in R-type MoS$_2$/WSe$_2$, as illustrated in (a), where the excitonic potential is tuned by an applied perpendicular electric field $\varepsilon$. The inset in each panel shows the potential profile along the high symmetry points (A-B-C-A) of the moir\'e supercell. For $\varepsilon=\varepsilon_{0}$, the excitonic potential exhibits the same value at regions $A$ and $B$, whereas for $\varepsilon=2\varepsilon_0$ ($\varepsilon=0$), $A$ ($B$) becomes higher in energy than $B$ ($A$).}
    \label{fig:Moire-Geometry}
\end{figure}

Since the Dirac equation predicts ZBW with amplitude of the order of the Compton wavelength ($10^{-2}\ \lyxmathsym{\AA}$) and a frequency of $\omega_{ZB}\approx10^{21}\ \text{Hz}$, which are not accessible with current experimental techniques~\cite{Rusin2009}, a direct experimental observation of this effect is challenging. Therefore, the past decades have seen an increasing interest in the ZBW phenomena in different systems, such as ultracold atoms~\citep{Vaishnav2008,Merkl2008}, semiconductors~\citep{Schliemann2005,Zawadzki2005,Schliemann2006,Rusin2007a,Schliemann2008,Biswas2014}, carbon nanotubes~\citep{Zawadzki2006}, topological insulators~\citep{Shi2013}, crystalline solids~\citep{Ferrari1990,Zawadzki2010} and other systems~\citep{Cannata1991,Vonsovskii1993,Lamata2007,Cunha2019}. In fact, ZBW has been experimentally detected using quantum simulations of the Dirac equation based on trapped ions~\citep{Gerritsma2010}, Bose--Einstein condensates~\citep{Wang2010,LeBlanc2013,Qu2013} and, most recently, an optical simulation~\citep{Silva2019}. Since the characteristic frequency of ZBW is determined by the energy gap between the (pseudo-)spin states~\cite{Zawadzki2005}, designing a system where the gap in the Dirac cone can be controlled at will is fundamental for optimization of the oscillation frequency and eventual experimental detection of this phenomenon.  


Most recently, advances in the isolation of monolayer semiconductors and their stacking as van der Waals heterostructures (vdWhs) opened a new field of study of artificial 2D hybrid materials~\cite{li2016heterostructures,liu2016van}. Combining two monolayers of semiconducting transition-metal dichalcogenides (TMDs) in a vdWhs with an inter-layer twist introduces an in-plane moir\'e pattern~\cite{zhang2017interlayer}, as illustrated in Fig.~\ref{fig:Moire-Geometry}(a). This pattern is associated with an in-plane modulation of the conduction and valence band edges, thus presenting new possibilities to engineer the electronic band structure, quasi-particle confinement, and optical properties of the system. Especially, inter-layer excitons (ILE) are profoundly affected by the moir\'e pattern, which creates regions in space where the ILE energy is significantly lower. For $\text{MoS}_{2}/\text{WSe}_{2}$ vdWhs with small twist angle (R-type stacking), lowest energy regions are those with stacking registry $R_{h}^h$ and $R_{h}^X$, represented by $A$ and $B$ in Fig.~\ref{fig:Moire-Geometry}(a). These regions form a honeycomb superlattice for excitonic confinement, thus leading to a moir\'e exciton band structure that resembles the one of gapped monolayer graphene. Different inter-layer distances for $R_{h}^h$ and $R_{h}^X$, as illustrated in Fig. \ref{fig:Moire-Geometry}(b), lead to different ILE dipole moments in each region. Consequently, a perpendicularly applied electric field $\varepsilon$ can be used to tune the energies of $A$ and $B$ ILE sub-lattices, thus making them equal at $\varepsilon = \varepsilon_0 \approx 0.44$ V/nm.\cite{yu2017moire} In this case, the moir\'e exciton band structure acquires a massless Dirac fermion character, as illustrated in Fig.~\ref{fig:Moire-Geometry}(c). The combination of the long lifetime and bright luminescence~\cite{yu2017moire} of ILE, along with their Dirac-like dispersion tunable by the twist angle and applied fields, makes twisted vdWhs a strong candidate for experimental detection of ZBW of moir\'e excitons.

In this letter, we analyze the dynamics of moir\'e exciton wave packets as an opto-electronics-based platform to probe ZBW, as an alternative to the previous proposals, mostly based on low-energy electrons in graphene or on ultra-cold atoms. To do so, we apply the time-evolution operator ~\cite{Chaves2015,Costa2015} on a wave packet distribution representing a moir\'e exciton in twisted $\text{MoS}_{2}/\text{WSe}_{2}$ vdWhs. We discuss the effects of the wave packet parameters, such as its pseudospinor and width, as well as of an applied electric field and different twist angles, on the ZBW amplitude and time evolution of the exciton probability density distribution. The optimization of parameters proposed here may guide future experiments towards the experimental observation of ZBW of such neutral quasi-particles in this vdWhs, which represents an important advance in the understanding not only of this phenomenon, but also of the tunable Dirac-like character of the moir\'e exciton.


\textit{Tight-binding approach for excitons in a potential landscape} ILEs in a twisted hetero-bilayer experience a periodic potential of the form~\cite{yu2017moire}
\begin{equation}
V\left(\vec{r}\right)=E_{g}\left(\vec{r}_{0}\left(\vec{r}\right)\right)+e\varepsilon d\left(\vec{r}_{0}\left(\vec{r}\right)\right)-E_{b}~,
\label{eq. potential moire}
\end{equation}
where $d$ is the inter-layer distance and $E_g$ the ILE bandgap, both modulated along the plane due to the moir\'e pattern (see Fig.~\ref{fig:Moire-Geometry}), and $\varepsilon$ is a perpendicularly applied electric field. Here, $\vec{r}_{0}$ is the in-plane displacement vector from a metal site in the hole layer to a nearest-neighbor metal site in the electron layer, depending on the location $\vec{r}$ in the moir\'e pattern. The binding energy $E_b$, on the other hand, is not expected to be significantly dependent on the local potentials~\cite{yu2017moire} and is, therefore, assumed to be constant. 

Excitons in such a potential landscape would be trapped at their local minima and exhibit a non-zero (complex) hopping to the neighboring minima. In a twisted $\text{MoS}_{2}/\text{WSe}_{2}$  bilayer, this landscape of energy minima has a honeycomb symmetry, with $A$ ($R_h^h$) and $B$ ($R_h^X$) sub-lattices at slightly different energies, $+\delta$ and $-\delta$, respectively. A low-energy quasi-particle - in this case, an exciton - in such a landscape would behave as a non-zero mass Dirac-Weyl fermion, whose Hamiltonian, within third-nearest neighbors approach, reads~\cite{ibanez2013tight,yu2017moire}
\begin{equation}
H_{mex}\hspace{-0.6mm}=\hspace{-0.6mm}\left( \begin{tabular}{cc}
$\delta - t_A F(\vec{k})$  & $t_0Z_0(\vec{k}) + t_2Z_2(\vec{k})$  \\
$t_0Z^*_0(\vec{k}) + t_2Z^*_2(\vec{k})$ & $-\delta - t_B F(\vec{k})$ \\
\end{tabular}  \right) ,
\label{Hamiltonian}
\end{equation}
where $t_{A(B)}$ is the hopping between nearest-neighbors minima of the $A$ and $B$ sub-lattices (see SM) that compose the honeycomb moir\'e potential, $t_0$ and $t_2$ are hopping parameters between first and third nearest-neighbors, see Fig.~\ref{fig:Moire-Geometry}(d), and structure factors are given by %
\begin{gather*}
F(\vec{k})=2\cos\left[\vec{k}\cdot(\vec{a_{1}}-\vec{a_{2}})-\theta_s\right]+ \hspace{2cm} \nonumber\\
\hspace{2cm}2\left[\cos(\vec{k}\cdot\vec{a_{1}}+\theta_s) +\cos(\vec{k}\cdot\vec{a_{2}}-\theta_s)\right],\nonumber \\
Z_{0}(\vec{k})=1+e^{-i(\vec{k}\cdot\vec{a_{1}}+\theta_s)}+e^{-i(\vec{k}\cdot\vec{a_{2}}-\theta_s)},\nonumber \\
Z_{2}(\vec{k})=e^{-i\vec{k}\cdot(\vec{a_{1}}+\vec{a_{2}})}+2\cos{[\vec{k}\cdot(\vec{a_{1}}-\vec{a_{2}})+\theta_s]},
\end{gather*}
where $\theta_s = 4\pi s/3$ originates from the complex part of the hopping parameters of the moir\'e exciton \cite{yu2017moire} with spin sign $s = \pm 1$.


Diagonalization of $H_{mex}$ leads to the moir\'e exciton band structure
\begin{equation}
E_{\pm} = -t_+ F(\vec{k}) \pm \sqrt{|t_0 Z_0(\vec{k}) + t_2Z_2(\vec{k})|^2 + (t_-F(\vec{k})-\delta)^2}~,
\end{equation}
where $t_{\pm} = (t_A \pm t_B) /2$. An example of such a band structure is shown in Fig.~\ref{fig:Moire-Geometry}(c). In the absence of external field, since the energies of sub-lattices $A$ and $B$ are different [see left panel in Fig.~\ref{fig:Moire-Geometry}(e)], $\delta \neq 0$ and the moir\'e exciton band structure exhibits a gap, as illustrated by solid lines in Fig. \ref{fig:Moire-Geometry}(c). However, as the applied field  $\varepsilon$ increases, the sub-lattices become similar in energy and $\delta$ approaches zero as the field reaches a critical value $\varepsilon_0$, which is 0.44 V/nm for the vdWhs considered here [see middle panel in Fig.~\ref{fig:Moire-Geometry}(e)]. In this case, the dashed lines in Fig.~\ref{fig:Moire-Geometry}(c) exhibit a gapless Dirac-like band structure for the moir\'e exciton in the vicinity of the $\Gamma$-point of the moir\'e Brillouin zone, which corresponds to either the K or K' points of the crystal Brillouin zone. Different colors in Fig. \ref{fig:Moire-Geometry}(c) stand for the excitonic band structures of the two possible exciton spins, up or down for $s=+$ (red) or $s=-$ (blue), respectively. Due to the spin-valley locking, the spin-valley index $s$ also corresponds to a moiré exciton at the \textbf{K} (\textbf{K'}) valley for $s=+$ ($-$) in the considered case of R-type stacking registry. As we will consider only large moir\'e exciton wave packets centered at $\Gamma$, where the bands for the two different spins are similar, spins are not expected to play a significant role in this study.  

\textit{Wave-packet dynamics} Writing the Hamiltonian as $H=\vec{\alpha}\cdot\vec{\sigma}$, where $\vec{\sigma}$ are the Pauli matrices, allows one to easily apply the time-evolution operator in an exact form as a simple matrix multiplication~\cite{Chaves2015,Costa2015,cunha2019wave}. Therefore, it is convenient to re-write Eq.~(\ref{Hamiltonian}) as
\begin{equation}
    H_{mex}=\vec{\alpha}(\vec{k})\cdot\vec{\sigma}-t_{+}F(\vec{k})\mathbb{1}~,
    \label{eq:new_hamiltonian}
\end{equation}
where $\mathbb{1}$ is the identity matrix and  $\vec{\alpha}=(\alpha_{x}(\vec{k}),-\alpha_{y}(\vec{k}),\alpha_{z}(\vec{k}))$ with its components given by
\begin{widetext}
\begin{subequations}\label{Eq. alpha of the hamiltonian}
%
%
%
%
%
%
%
\begin{gather}
\alpha_{x}\qty(\bf{k})=\qty[1+\text{cos}(\theta{_s}+\vec{k}\cdot\vec{a}_{1})+\text{cos}(\theta_s-\vec{k}\cdot\vec{a}_{2})]t_{0}
+\left\{\text{cos}\left[(\vec{a}_{1}+\vec{a}_{2})\cdot\vec{k}\right]+2\text{cos}\left[\theta_s+(\vec{a}_{1}-\vec{a}_{2})\cdot\vec{k}\right]\right\}t_{2},    \tag{\theequation a} \\
\hspace{1cm}\alpha_{y}(\vec{k})=[\text{sin}(\theta_{s}+\vec{k}\cdot\vec{a_{1}})-\text{sin}(\theta_{s}-\vec{k}\cdot\vec{a_{2}})]t_{0}+\text{sin}\left[(\vec{a_{1}}+\vec{a_{2}})\cdot\vec{k}\right]t_{2} \quad
\text{and}  \quad
\alpha_{z}(\vec{k})=\delta-t_{-}F(\vec{k})~. \tag{\theequation b,c}
\end{gather}
\end{subequations}
\end{widetext}
Since $H_{mex}$ does not explicitly depend on time and $\left[\vec{\alpha}\cdot\vec{\sigma},-t_{+}F(\vec{k})\mathbb{1}\right]=0$, the time-evolution operator for the Hamiltonian defined in Eq.~(\ref{eq:new_hamiltonian}) is given by
\begin{equation}
e^{-\frac{i}{\hbar}H_{mex}\Delta t}=e^{-i\vec{\beta}\cdot\vec{\sigma}}e^{-\frac{i}{\hbar}(-t_{+}F(\vec{k})\mathbb{1})\Delta t}~,
\label{eq: split-operator}
\end{equation}
where $\vec{\beta}=\vec{\alpha}\Delta t/\hbar$.

From the well known properties of the Pauli matrices, the first exponential on the right hand-side of Eq.~(\ref{eq: split-operator}) yields
\begin{equation}
    e^{-i\vec{\beta}\cdot\vec{\sigma}}=\text{cos}\left(\beta\right)\mathbb{1}-\frac{i\text{sin}\left(\beta\right)}{\beta}\left(\begin{array}{cc}\beta_{z} & \beta_{x}-i\beta_{y}\\
    \beta_{x}+i\beta_{y} & \beta_{z}\end{array}\right)=\mathcal{M}~,
\label{eq: evolution_beta_sigma}
\end{equation}
where $\beta=|\vec{\beta}|$, and the second exponential of Eq.~(\ref{eq: split-operator}) is equivalent to
\begin{equation}
e^{\frac{i}{\hbar}(t_{+}F(\vec{k})\mathbb{1}\Delta t)}=\mathbb{1}e^{\frac{i}{\hbar}(t_{+}F(\vec{k})\Delta t)}=\mathcal{N}~.\label{eq: exponential_second_term}
\end{equation}
Applying the time-evolution operator defined in Eq.~(\ref{eq: split-operator}) on the wave function $\Psi\left(\vec{r},t\right)$, one obtains the propagated wave function after a time step $\Delta t$ as
\begin{equation}
\Psi\left(\vec{r},t+\Delta t\right)=e^{-\frac{i}{\hbar}H_{mex}\Delta t}\Psi\left(\vec{r},t\right)=\mathcal{M}\mathcal{N}\Psi\left(\vec{r},t\right)~.
\label{time-evolution operator}
\end{equation}
Note that $\mathcal{M}$ and $\mathcal{N}$ depend on the wave vector $\vec{k}$, therefore, the matrix multiplication with a general initial wave packet is conveniently computed numerically in reciprocal space by performing a Fourier transform on the wave function, which gives this method a flavor of a semi-analytical procedure. At $t=0$, we assume the wave function as a circularly-symmetric 2D Gaussian wave packet with width $d$ multiplied by the pseudospinor $[C_1~C_2]^{T}$, such as
\begin{equation}
\Psi\hspace{-0.7mm}\left(\vec{r},t\right)\hspace{-0.7mm}=\hspace{-0.7mm}N\left(\begin{array}{c} C_1 \\ C_2\end{array}\right)\hspace{-0.2mm}\text{exp}\hspace{-0.7mm}\left[\hspace{-0.7mm}-\frac{\left(x-x_{0}\right)^{2}-\left(y-y_{0}\right)^{2}}{d^{2}}\right],
\label{eq: wave function t=0}
\end{equation}
where $N$ is the normalization factor and $(x_0,y_0)$ are the coordinates of the center of the Gaussian wave packet in real space. As the exciton is normally excited by a low-momentum photon, we assume a moir\'e exciton exactly at the $\Gamma$-point of the moir\'e Brillouin zone, i.e. with zero energy and zero momentum.  

\begin{figure}[!t]
\centering{\includegraphics[width=0.7\columnwidth]{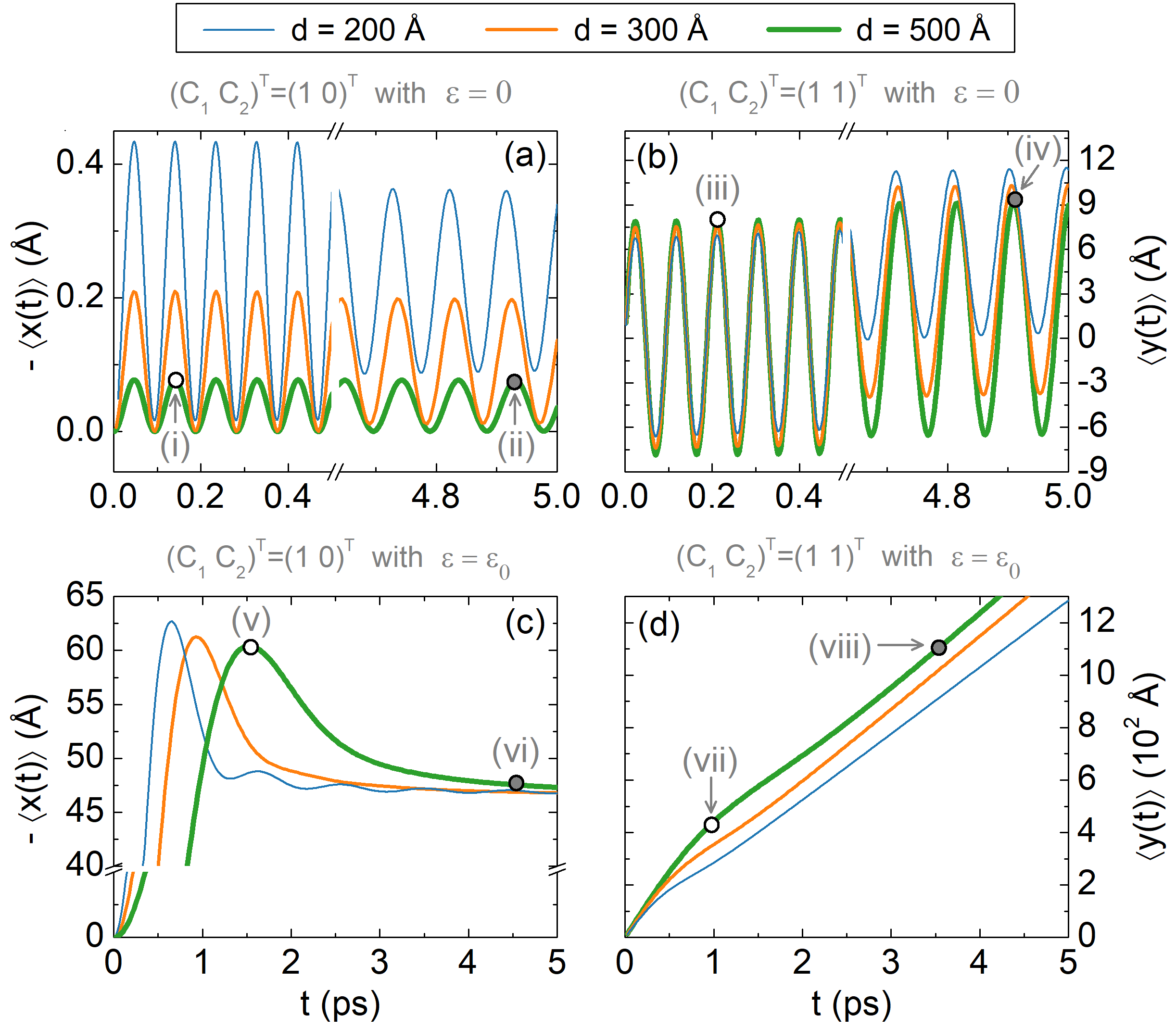}}
    \caption{(Color online) ZBW of the expectation values of the position of a moir\'e exciton in a MoS$_2$/WSe$_2$ vdWhs, considering an initial Gaussian wave packet distribution with $d=200~\text{\AA}$ (blue), $d=300~\text{\AA}$ (orange) and $d=500~\text{\AA}$ (green), and pseudo-spinors $[\text{C}_1~\text{C}_2]^{\text{T}}=[1~0]^{\text{T}}$ and $[\text{C}_1~\text{C}_2]^T=[1~1]^T$, under applied fields (a,b) $\varepsilon = 0$ and (c,d) $\varepsilon = \varepsilon_0$. The propagated probability densities for the time instants marked with white and gray circular dots in each panel are shown in Fig.~\ref{fig: Wave Function snap shoot}.}
    \label{fig:ZBW_averages_ta_dif_tb_all}
\end{figure}

\textit{Wave packet dynamics and zitterbewegung} Figure \ref{fig:ZBW_averages_ta_dif_tb_all} illustrates the average position $\expval{x\qty(t)}$ and $\expval{y\qty(t)}$ of the wave packet as a function of time for $d=200~\text{Å}$ (blue), $300~\text{Å}$ (orange) and $d=500~\text{Å}$ (green). Different pseudo-spin polarizations $\qty[C_1~C_2]^\text{T}=\qty[0~1]^\text{T}$ and $\qty[1~1]^\text{T}$ are considered, with and without an applied electric field $\varepsilon$, as indicated on top of each panel. Results for $\qty[1~i]^\text{T}$ are given in the Supplemental Material (SM), along with the material parameters for the vdWhs studied here. The pseudo-spinor represents the occupation of the $A$ and $B$ sub-lattice sites, therefore, it is expected to be controlled in an actual experiment by the polarization of the excitation light, since the $R_h^h$ and $R_h^X$ regions, which correspond to the $A$ and $B$ sub-lattices here, exhibit different selection rules for circular light polarization~\cite{yu2017moire}. For instance, a circular light polarization that excites ILE only in $R_h^h$ ($R_h^X$) regions would effectively produces a moir\'e exciton wave packet with pseudo-spinor $\qty[C_1~C_2]^\text{T}=\qty[1~0]^\text{T}$ ($\qty[0~1]^\text{T}$). As for the wave packet width, it could be controlled e.g. by the focus of the short-pulse excitation light, although actual precise manipulation and engineering of excitonic wave packets may be a challenging task~\cite{zang2017engineering}. Laser spots as narrow as $\approx$ 500 \AA\, i.e. of the same order of magnitude as the wave packets considered here, have been used for the study of exciton dynamics in 2D semiconductors in recent experiments~\cite{zipfel2020exciton, perea2019exciton, kulig2018exciton, unuchek2019valley}.

In the absence of an external applied electric field ($\varepsilon=0$), both expectation values $\expval{x\qty(t)}$ and $\expval{y\qty(t)}$ exhibit ZBW, but with very low amplitude and high frequency, which hinders the actual observation of this effect. On the other hand, for $\varepsilon=\varepsilon_0$, where the gap is closed and moir\'e exciton effectively behaves as a massless Dirac quasi-particle, the wave packet moves only in one direction, exhibiting damped oscillations. For conciseness, Figs. \ref{fig:ZBW_averages_ta_dif_tb_all} (c,d) show only the moving component of $\vec{r}$, see SM for the other component. In this case, the amplitude of the oscillation is much higher, of the order of tens of \AA\, with a timescale of the order of few pico-seconds, which would make this effect clearly observable in actual experiments. Wave packets with smaller width exhibit weak oscillations, which vanish as the width increases. Nevertheless, for a $[1~0]^T$ spinor wave packet, a $\approx 60$ \AA\, peak, followed by a $\approx 50$ \AA\, permanent shift of the center of the wave packet, is observed for all values of wave-packet width considered here. For larger widths, the motion resembles the one of zero-energy electron wave packets in monolayer graphene~\cite{lavor2020effect, Maksimova2008}, since the wave packet becomes narrower around the $\Gamma$-point of the moir\'e Brillouin zone, where dispersion is approximately the same as in graphene. The dependence of the maximum displacement of the expectation value $\expval{x(t)}$ as a function of $\epsilon$, as well as the time for this maximum displacement to occur, is discussed in the SM, where it is demonstrated that both the maximum wave packet displacement and its timescale are highest at $\varepsilon = \varepsilon_0$. 

As for a $[1~ 1]^T$ spinor wave packet, the center of mass is predicted to move almost linearly with time, travelling tens of \AA\, in just a few picoseconds, before the exciton recombines.  
\begin{figure}[!t]
\centering{\includegraphics[width=0.7\columnwidth]{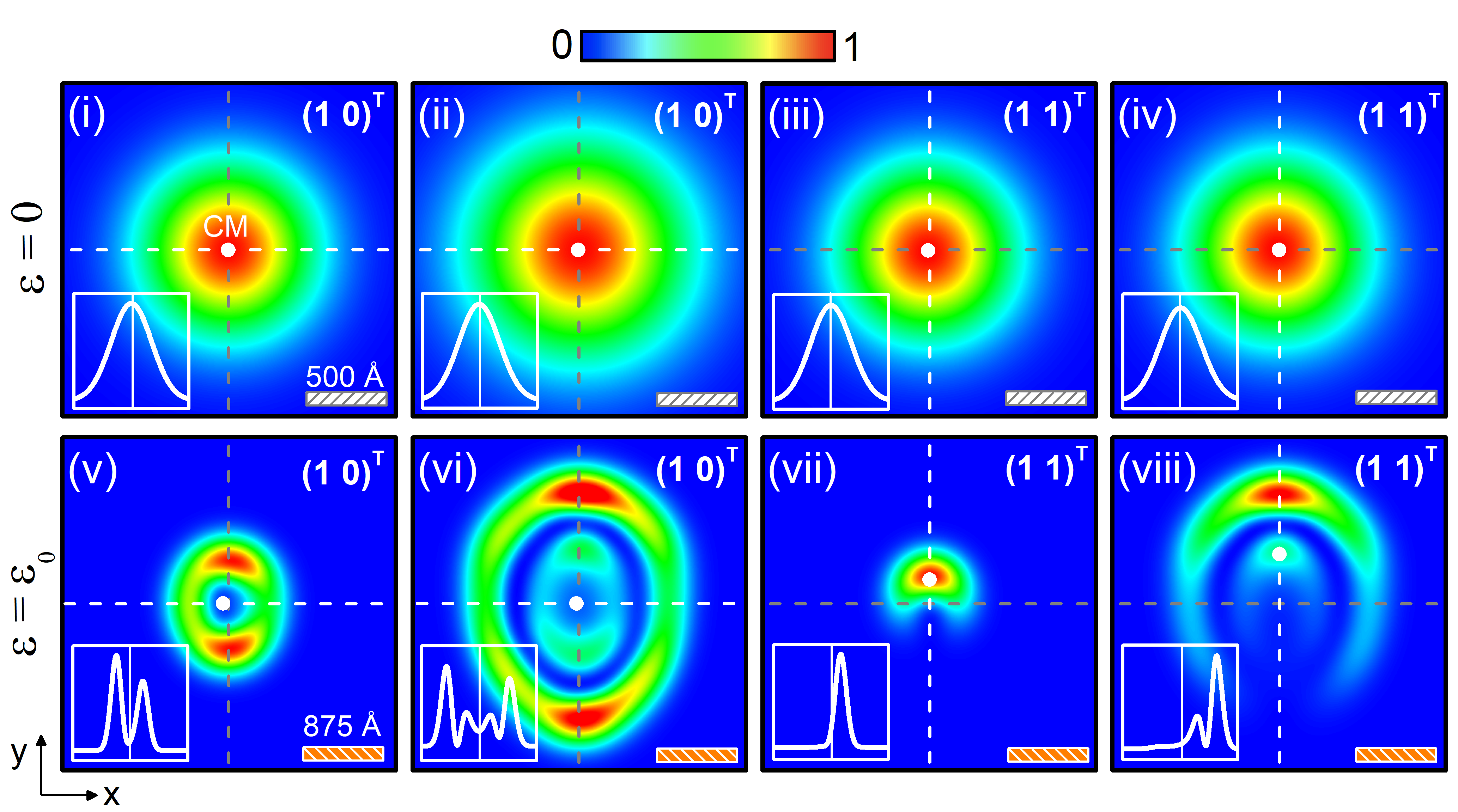}}
    \caption{(Color online) Snapshots of the propagated probability density $\qty|{\Psi\qty({\vec{r}},t)}|^{2}$ for an initial Gaussian wave packet with width $d=500~\text{Å}$ and pseudo-spinors $\qty[1~0]^T$ and $\qty[1~1]^T$. Top (bottom) row shows results for applied electric field $\varepsilon = 0$ ($\varepsilon = \varepsilon_0$). The white (orange) bar corresponds to $500~${\AA} ($875~${\AA}) and the small white dot inside each panel represents the center-of-mass of the wave packet. The profiles of $\qty|{\Psi\qty({\vec{r}},t)}|^{2}$ along the dashed white lines in each panel are shown as insets. The labels (i) to (viii) correspond to different time steps as marked with circular dots in Fig.~\ref{fig:ZBW_averages_ta_dif_tb_all}.}
    \label{fig: Wave Function snap shoot}
\end{figure}

The unique features predicted here for the moir\'e exciton wave packet dynamics can also be observed in the probability density distribution, as shown in Fig. \ref{fig: Wave Function snap shoot}. An initial Gaussian wave packet for the center-of-mass coordinate of a free exciton is expected to simply disperse across space as time elapses. Similar dispersion is observed e.g. in Refs.~[\onlinecite{zipfel2020exciton}] and [\onlinecite{kulig2018exciton}] for monolayer TMD. Notice, however, that the Gaussian packet in those experiments represented a density distribution of multiple excitons, rather than an actual single exciton wave function, so that phonon effects\cite{glazov2019phonon}, which give rise to a halo in the exciton distribution, play an important role. In our case, in order to avoid such phonon hot-spots \cite{glazov2019phonon} and exciton-exciton interaction effects, one would have to employ low intensity exciting irradiation at low temperatures, so that the exciton distribution effectively matches the non-interacting excitons picture proposed here. In this case, the moir\'e exciton wave packet evolves as a double ring structure in the presence of an electric field with the critical value $\varepsilon_0$, whereas the usual dispersion is observed in the absence of field. The observation of this strikingly different wave packet dispersion in time in the presence of the applied field would thus represent a smoking gun evidence of the ZBW of moir\'e excitons and their Dirac-like nature.  

%
%
%
%
%



\textit{Conclusion} In summary, we argue that dynamics of a moir\'e exciton wave packet is an advantageous solid-state opto-electronic platform to probe ZBW, evasive in experiments to date. In MoS$_2$/WSe$_2$ vdWhs with small twist angles, the moir\'e pattern created by the inter-layer lattice mismatch produces a periodic in-plane potential for the ILE center-of-mass and, consequently, a moir\'e exciton band structure. A moir\'e exciton wave packet in this system exhibits very fast and weak oscillations, hard to detect experimentally. However, in the presence of a perpendicular electric field, the gap of the moir\'e exciton band structure can be closed, which attributes the characteristics of a massless Dirac fermion to this quasi-particle, so ZBW becomes naturally more evident. In such a case, we reveal a shift of tens of \AA\, in the center of the moir\'e exciton wave packet, along with damped oscillations with pico-second long periods. The exciton probability density profile is demonstrated to be strikingly different in the presence of gap-closing electric field, compared to the case without any field. The density profile and motion is also shown to be strongly dependent on the pseudo-spinor of the moir\'e exciton wave packet, which is controllable by the polarization of the incident exciting light. With relevant timescales being within reach of available experimental techniques, we expect to instigate the first experimental detection of ZBW in an exciton wave packet, which opens the gate to follow-up studies exploiting thereby proven massless Dirac fermion character of the moir\'e excitons in MoS$_2$/WSe$_2$ vdWhs induced by gating. 

\textit{Acknowledgements} This work was supported by the Brazilian Council for Research (CNPq), through the PRONEX/FUNCAP, Universal, and PQ programs, the Brazilian National Council for the Improvement of Higher Education (CAPES), and the Research Foundation - Flanders (FWO).

\bibliographystyle{apsrev4-2}
\bibliography{myreferences}

\newpage

\section*{\Large Supplemental Material for\\ ``Zitterbewegung of moir\'e excitons in twisted MoS$_2$/WSe$_2$ hetero-bilayers''}

In this Supplemental Material file, we present (i) the parameters of the moiré exciton used in the main text; (ii) the expression for the hopping energy dependence on the moiré trapping potential and other system parameters; and (iii) the expectation values of the position $\expval{\Vec{r}\qty(t)}$ of a moiré exciton in a MoS2/WSe2 van der Waals heterostructure (vdWhs) for an initial Gaussian wave packet with different pseudo-spinors, in addition to those discussed in the main text.

\section{Material parameters of R-type $\text{MoS}_2\text{/WSe}_2$: inter-layer exciton bandgap and moiré exciton band structure}

\begin{figure}[H]
\centering{\includegraphics[width=0.43\columnwidth]{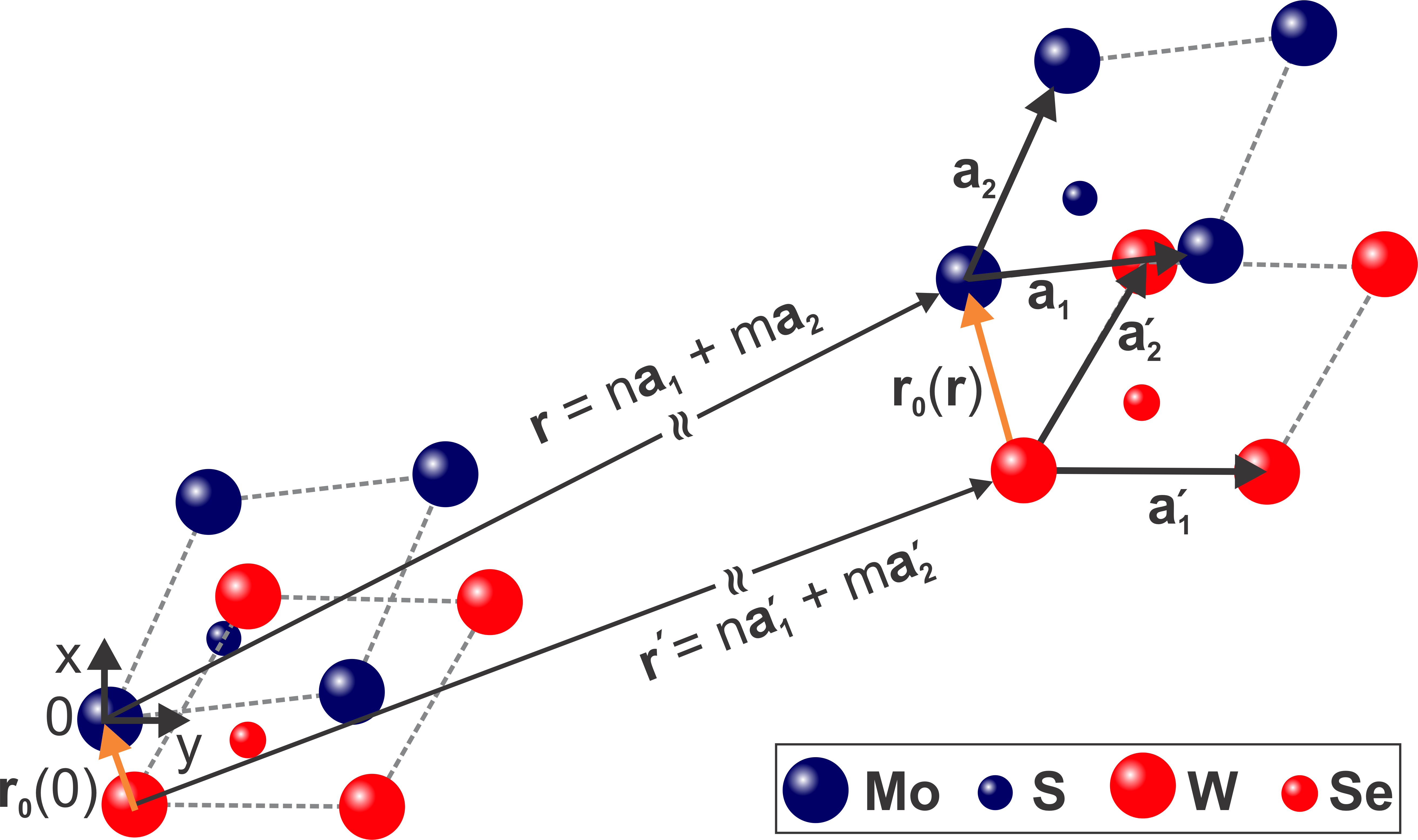}}
    \caption{(Color online) Dependence of interlayer translation vector $\Vec{r}_0(\Vec{r})$ on the interplane position vector $\Vec{r}$ in a MoS$_2$/WSe$_2$ hetero-bilayer (adapted from Ref.~\cite{yu2017moire}).}
    \label{fig:Complementary_results_hetero-bilayer}
\end{figure}

An important consequence of the moiré pattern in a twisted MoS$_2$/WSe$_2$ hetero-bilayer is the fact that the inter-layer excitons bandgap, $E_g(\Vec{r}_0)$, is a function of the in-plane displacement vector from a metal site in the hole layer to a nearest-neighbor metal site in the electron layer. In turn, $\Vec{r}_0(\Vec{r})$ depends on the location $\Vec{r}$ in the moiré pattern. A complete description of the approximation to obtain the equation for $E_g(\Vec{r}_0)$ can be found in Ref.~\cite{wang2017interlayer} and also in the Supplementary Material of Ref.~\cite{yu2017moire}. Therefore, we will limit ourselves here to just reproducing such equation, for the sake of completeness, which is defined as:

\begin{equation}
    \tag{S1}
    E_g(\Vec{r}_0) = E_{g,0} + {\Delta}E_{g,1}\abs{f_0(\Vec{r_0})}^2 + {\Delta}E_{g,2}\abs{f_+(\Vec{r_0})}^2~,
    \label{ILE_bandgap}
\end{equation}
with the mapping from the moiré supercell to the monolayer unit cell defined by the function
\begin{equation}
    \tag{S2}
    \Vec{r}_0(\Vec{r}) = \Vec{r}_0(0) + \Vec{R}-\Vec{R'} = \Vec{r}_0(0) + n(\Vec{a}_1 - \Vec{a}_{1}') + m(\Vec{a}_2 - \Vec{a}_{2}')~,
\end{equation}
where $\Vec{r}\equiv n\Vec{a}_1 + m\Vec{a}_2$ and $\Vec{r}'\equiv n\Vec{a}_1' + m\Vec{a}_2'=(1+\delta)\hat{C}_{-\delta\theta}\Vec{r}$. The primitive lattice vectors of WSe$_2$ (MoS$_2$) are given by $\Vec{a}_{1,2}'$ ($\Vec{a}_{1,2}=\frac{1}{1+\delta}\hat{C}_{\delta\theta}\Vec{a}_{1,2}'$), see Fig.~\ref{fig:Complementary_results_hetero-bilayer}. $n$ and $m$ are integers and $\hat{C}_{-\delta\theta}$ represents the rotation of $\Vec{r}$ by an angle $-\delta\theta$. In Eq.~(\ref{ILE_bandgap}), as discussed in Ref.~\cite{wang2017interlayer}, $\Vec{f}_0(\Vec{r}_0)$ and $\Vec{f}_\pm(\Vec{r}_0)$ are defined, respectively, as:
\begin{equation}
    \tag{S3}
    f_0(\Vec{r}_0) =  \frac{e^{-i\Vec{K}\cdot\Vec{r}_0} + e^{-i\hat{C_3}\Vec{K}\cdot\Vec{r}_0} + e^{-i\hat{C_{3}^{2}}\Vec{K}\cdot\Vec{r}_0}}{3}~,
\end{equation}
and
\begin{equation}
    \tag{S4}
    f_{\pm }(\Vec{r}_0) =  \frac{e^{-i\Vec{K}\cdot\Vec{r}_0} + e^{-i(\hat{C_3}\Vec{K}\cdot\Vec{r}_0 \pm \frac{2\pi}{3})} + e^{-i(\hat{C_{3}^2}\Vec{K}\cdot\Vec{r}_0 \pm \frac{4\pi}{3})}}{3}~.
\end{equation}

\noindent
The coupling between two bands in different layers at the $\Vec{K}$-point, considering only the leading Fourier components, are defined by $\Vec{f}_\pm(\Vec{r}_0)$, as discussed in Refs.~\cite{wang2017interlayer, yu2017moire}.

On the other hand, the interlayer separation can also be defined as~\cite{yu2017moire}
\begin{equation}
    \tag{S5}
    d(\Vec{r}_0) = \Vec{d}_0 + {\Delta}d_{1}\abs{f_0(\Vec{r_0})}^2 + {\Delta}d_{2}\abs{f_+(\Vec{r_0})}^2~,
\end{equation}
obtained from an experimental data fitting in Ref.~\cite{yu2017moire} (for more details, see Ref.~\cite{yu2017moire}, Sec.~II).

All parameters used to obtain the moiré exciton band structure and the colormap of the inter-layer exciton in Figs.~1(c) and ~1(e) in the main text, respectively, are summarized in Tab.~\ref{Tab: moire_band_structure}.
%

\begin{table}[H]
\centering{}
\caption{The parameters to obtain the colormap and moiré exciton band structure of R-type MoS$_2$/WSe$_2$ hetero-bilayer obtained from Refs.~\cite{wang2017interlayer,yu2017moire}.}
\begin{tabular}{| c | c | c |}
\hline\hline
Variable                    && Value     \\
\hline
$b$                         &&  10~nm     \\
$\delta$                    &&  3~meV     \\
$\Delta E_{g,1}$            &&  -116~meV     \\
$\Delta E_{g,2}$            &&  -94~meV     \\
$d_{0}$                     && 6.387~\AA     \\
$\Delta d_{1}$              &&    0.544~\AA     \\
$\Delta d_{2}$              &&    0.042~\AA     \\
\hline\hline
\end{tabular}
\label{Tab: moire_band_structure}
\end{table}

\section{Hopping strength of the exciton bands in a superlattice potential}

The hopping magnitude $t_\nu$ ($\nu=A,B$) between nearest-neighbors minima of the $A$ and $B$ sub-lattices, used in Eq.~(2) in the main text, can be approximated as \cite{yu2017moire,ibanez2013tight}
\begin{equation}
    \tag{S6}
    t_\nu\approx 0.78 E_R(V_{\nu}\qty(\varepsilon)/E_R)^{1.85}\text{exp}\qty[-3.404\sqrt{V_{\nu}\qty(\varepsilon)/E_R}]~,
\end{equation}
where $E_R=\tfrac{\hbar^2}{2M_0}\left ( \frac{4\pi}{3b} \right )^2$ is the recoil energy with $M_0=0.8m$ \cite{zhang2020twist} being the exciton mass written in units of the free-electron mass $m$, and $V_{\nu}\qty(\varepsilon)$ is the confining barrier height of $A$ and $B$ minima, as defined in Fig.~1 in the main text.

\section{Wave packet dynamics and zitterbewegung: complementary results}
In this section, we present complementary results of the average positions $\expval{x\qty(t)}$ and $\expval{y\qty(t)}$ of the Gaussian wave packet, not shown in Fig.~2 of the main manuscript, for the pseudo-spinors $[1~1]^T$ and $[1~0]^T$, respectively, as well as results for the $[1~i]^T$ pseudo-spinor.

\begin{figure}[H]
\centering{\includegraphics[width=0.6\columnwidth]{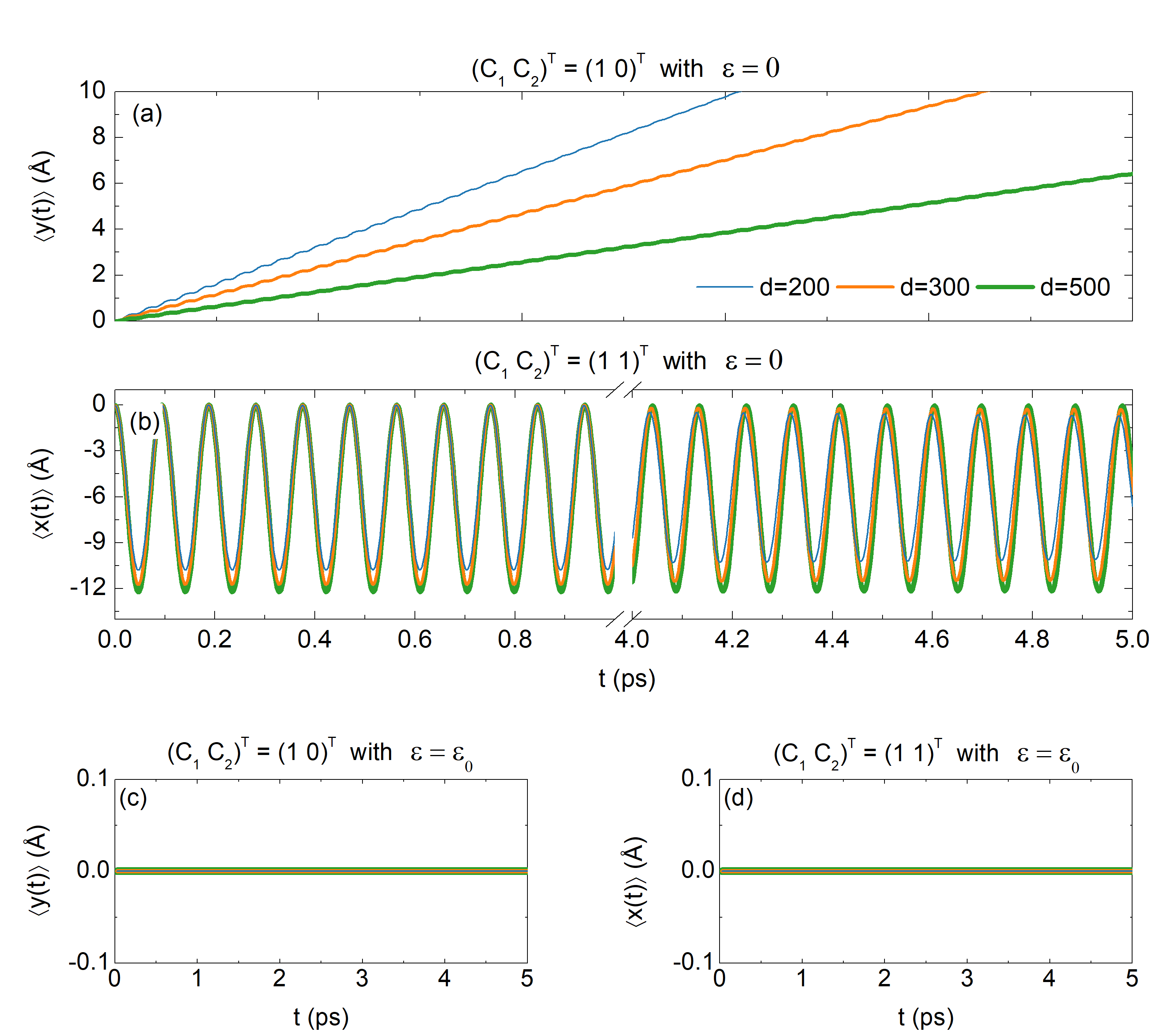}}
    \caption{(Color online) (a)-(d) \textit{Zitterbewegung} of a moiré exciton in a MoS$_2$/WSe$_2$ hetero-bilayer for an initial Gaussian wave packet distribution with $d=200~{\text{\AA}}$ (blue), 300~{\AA} (orange) and 500~{\AA} (green) and pseudo-spinors $[\text{C}_1~\text{C}_2]^{\text{T}}=[1~0]^{\text{T}}$ and $[\text{C}_1~\text{C}_2]^T=[1~1]^T$, under applied fields (a,b) $\varepsilon = 0$ and (c,d) $\varepsilon = \varepsilon_0$.}
    \label{fig:Complementary_results_1}
\end{figure}

Figure~\ref{fig:Complementary_results_1} presents results of the ZBW on the expectation values of the position of a moir\'e exciton in a MoS$_2$/WSe$_2$ vdWhs, considering an initial Gaussian wave packet distribution with width given by $d=200~\text{\AA}$ (blue), $d=300~\text{\AA}$ (orange) and $d=500~\text{\AA}$ (green), and pseudo-spinors $[\text{C}_1~\text{C}_2]^{\text{T}}=[1~0]^{\text{T}}$ and $[\text{C}_1~\text{C}_2]^T=[1~1]^T$, under applied fields (a,b) $\varepsilon = 0$ and (c,d) $\varepsilon = \varepsilon_0$. For both pseudo-spin configurations, when $\varepsilon=\varepsilon_0$, Figs.~\ref{fig:Complementary_results_1}(c,d) show that the ZBW is suppressed for $\expval{y\qty(t)}$ and $\expval{x\qty(t)}$ with $[1~0]^{\text{T}}$ and $[1~1]^T$, respectively. On the other hand, as shown in Fig.~2(c,d) in the main manuscript, $\expval{x\qty(t)}$ and $\expval{y\qty(t)}$ oscillate. Consequently, applying a perpendicular electric field ($\epsilon$) to the hetero-bilayer structure and considering the pseudo-spinor given by $[1~0]^{\text{T}}$ or $[1~1]^T$, means to restrict the wave packet propagation to only one direction in the $xy-$plane. If the electric field is zero, both coordinates of the center mass will exhibit ZBW, as one verifies in Fig.~\ref{fig:Complementary_results_1}(a,b) here and Fig.~2(a,b) in the main manuscript.

\begin{figure}[H]
\centering{\includegraphics[width=0.8\columnwidth]{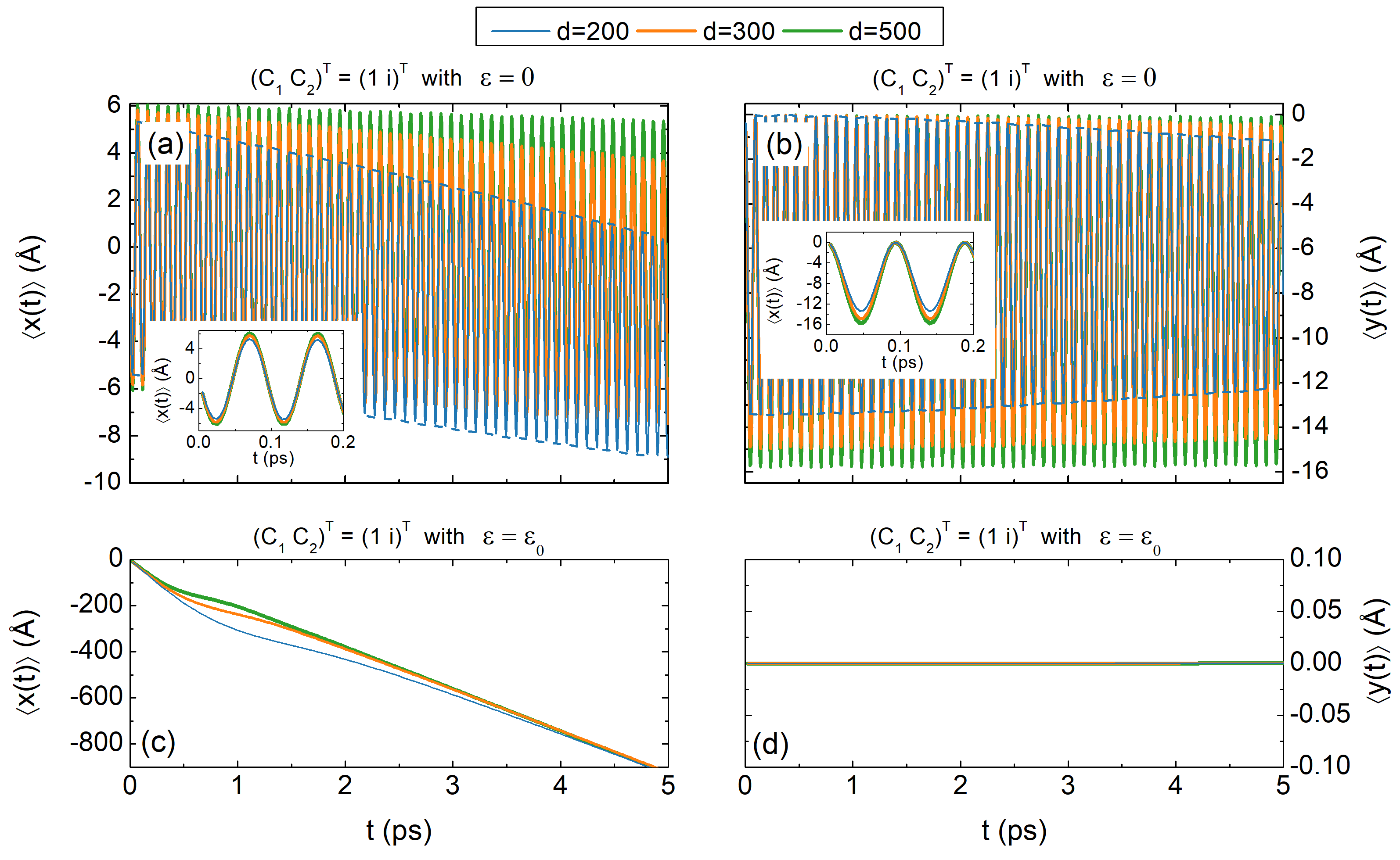}}
    \caption{(Color online) The same as in Fig.~\ref{fig:Complementary_results_1}, but now for a pseudo-spinor defined as $[\text{C}_1~\text{C}_2]^{\text{T}}=[1~i]^{\text{T}}$.}
    \label{fig:Complementary_results_2}
\end{figure}

As another example, we analyze a very commonly investigated initial pseudo-spinor polarization,  $[\text{C}_1~\text{C}_2]^{\text{T}}=[1~i]^{\text{T}}$. The ZBW for both coordinates of the center-of-mass of the Gaussian wave packet with ($\epsilon\neq0$) and without ($\epsilon=0$) an applied electric field $\epsilon$ are presented in Fig.~\ref{fig:Complementary_results_2} for different values of the packet width $d$. Similarly to the $[1~0]^T$ and $[1~1]^T$ pseudo-spinor configurations, when $\epsilon\neq0$, Fig.~\ref{fig:Complementary_results_2}(c,d), only one of the components features ZBW, i.e $\expval{x\qty(t)}\neq0$ and $\expval{y\qty(t)}=0$. On the other hand, for $\epsilon=0$, Fig.~\ref{fig:Complementary_results_2}(a,b), both coordinates oscillate, exhibiting ZBW with the same frequency for different width $d$ and a small difference in their amplitudes as time increases.

\vspace{1cm}

\begin{figure}[!t]
\centering{\includegraphics[width=0.5\columnwidth]{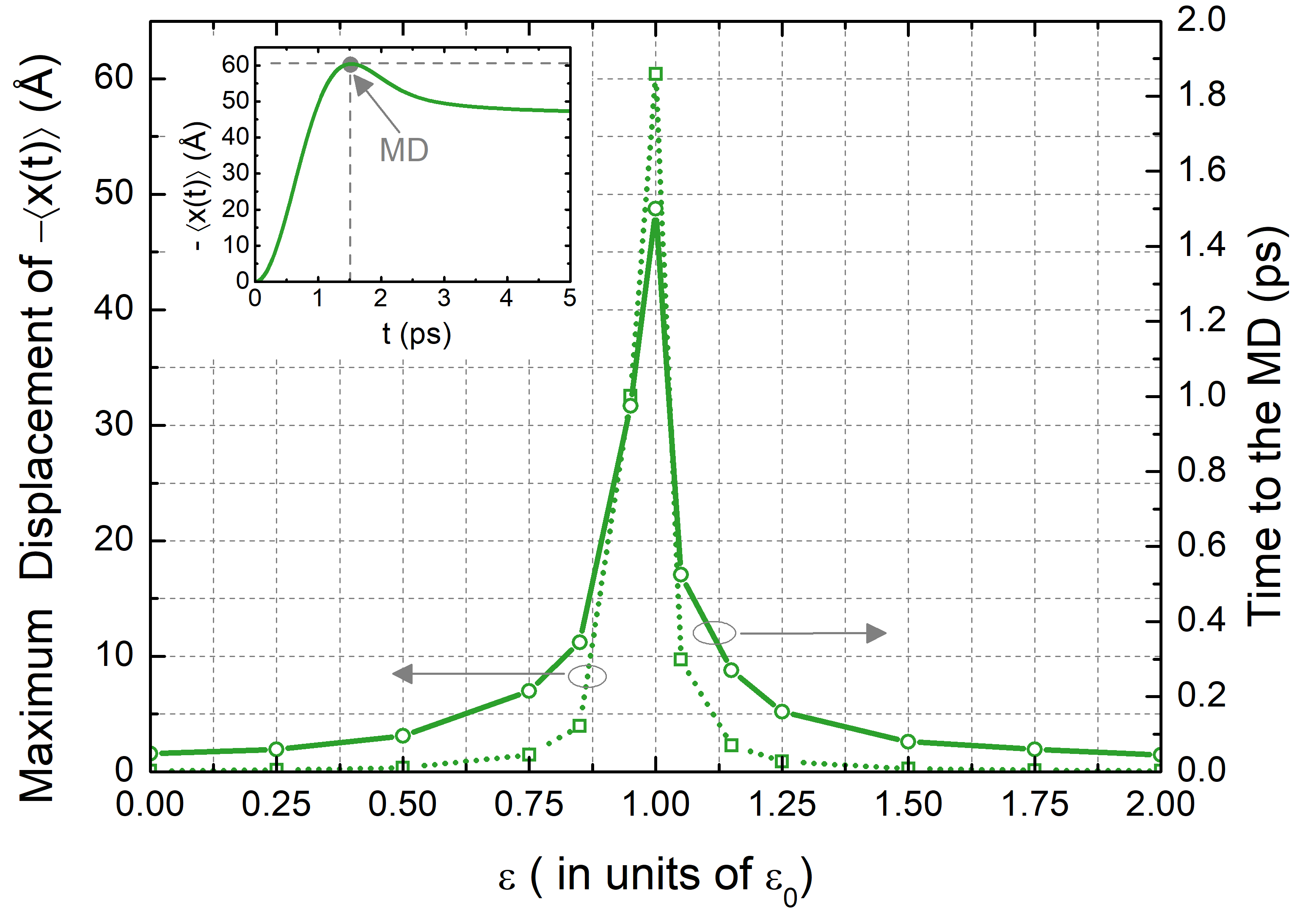}}
    \caption{(Color online) Dependence of the absolute value of the maximum displacement (MD) for the expectation value $\expval{x(t)}$ (left vertical axis), as well as the time to the MD, (right vertical axis), for an initial wave packet with $d = 500 \text{\AA}$ and pseudo-spinor $[\text{C}_1~\text{C}_2]^{\text{T}}=[1~0]^{\text{T}}$. For $\epsilon = \epsilon_0$, where $\epsilon_0 \approx 0.44 \text{V/nm}$, both results are maximized. The inset shows $\expval{x(t)}$ as a function of time at the critical field, see Fig.~2(c) in the main manuscript. The arrow indicates the time and magnitude of the MD.}
    \label{fig:Complementary_results_3}
\end{figure}

Finally, in order to reinforce the importance of the proper choice of the external electric field for the observation of the phenomena discussed in the main manuscript, the dependence of the absolute value of the maximum displacement (MD) for the expectation value $\expval{x(t)}$ of the wave packet as a function of the applied field is illustrated in Fig.~\ref{fig:Complementary_results_3}. Results are shown for a wave packet with $d=500 \text{\AA}$ width and pseudo-spinor $[\text{C}_1~\text{C}_2]^{\text{T}}=[1~0]^{\text{T}}$. When the applied electric field is $\epsilon = \epsilon_0$, where $\epsilon_0 \approx 0.44 \text{V/nm}$, both the MD and the time the wave packet takes to reach the MD (see right axis) are maximized. The inset in Fig.~\ref{fig:Complementary_results_3} shows the time dependence of $\expval{x(t)}$ at the critical field $\epsilon_0$, where the arrow identifies the time and magnitude of the MD. 

\end{document}